# Dark Energy may link the numbers of Rees

C Sivaram

Indian Institute of Astrophysics, Bangalore 560034, India

**Abstract:** There is increasing evidence that the universe is dominated by dark energy of the type given by an invariant cosmological constant. Latest data also indicates that fundamental interaction couplings and particle masses have remained remarkably constant from the earliest epochs. It is natural to connect these two 'steady state' features of the evolving universe, suggesting a role for the cosmic vacuum energy in fixing these interaction constants. Advances in high precision cosmology have revealed that dark matter of an unknown type constitutes about one-fourth of cosmic matter while baryons account for just four percent. These various cosmic parameters are enumerated by the six numbers of Rees. With the dark energy as a unifying link these numbers can all be connected and their values estimated.

There is increasing evidence that most of the energy density of the universe consists of a dark energy component with negative pressure that causes the cosmic expansion to accelerate. This is suggested by observation of large-scale structure; searches for type Ia supernovae at high redshifts and measurements of the cosmic microwave background anisotropy. [1] [2-5] Detailed WMAP measurements as well as the various supernovae-cosmology projects and the Hubble telescope have ushered in the era of high precision cosmology. The Hubble constant is now known to a few percent and the material inventory of the universe has been fixed. Dark Energy (DE) constitutes at least seventy percent of the universe, 26 percent is Dark Matter (DM) and about four percent is in the form of familiar baryonic matter. [6, 7]

Very recent work [8, 9] suggests that DE density is a constant, consistent with the cosmological constant first introduced by Einstein! [8] Recent indications that the dark energy is perhaps just the cosmological constant come from the Chandra Observatory which has made x-ray observations of hot gas in about 26 clusters of galaxies. [8] The supernova legacy survey (SNLS) is on track to detect hundreds of type Ia supernovae billions of light years distant. The first year of SNLS data turned up 71 type Ia supernovae.

By combining information on these, with data from the Sloan digital sky survey (SDSS), it appears that the case has been much strengthened [10] for DE being just Einstein's cosmological constant, implying a vacuum energy density that remains unchanged throughout space and time.

It is thus remarkable that in an evolving universe, the cosmological constant (appearing as a fundamental parameter), setting a cosmic scale, introduces a steady state feature (at least asymptotically!). It is equally remarkable that the coupling constants of various fundamental interactions and masses of elementary particles have remained invariant for the entire Hubble age! There were recent claims that spectral observations of distant quasars implied increase of the electromagnetic fine structure constant, $\alpha$, with epoch. [11] However, the latest results imply a zero time variation. [12] Thus, the remarkable constancy of the fundamental constants and particle masses is another 'steady state' feature of the evolving universe. This indicates that it may be natural to link the cosmic vacuum energy to local physical parameters like elementary particle masses and coupling constants. [13, 14] Some time back, Martin Rees in his book [15] enumerated six numbers needed to fix the universe. Apart from the vacuum energy cosmic term (there is increasing evidence, as stated above, that this is just Einstein's cosmological constant), the other numbers which are crucial for the evolution of the universe and for living systems are: (1) the binding energy of the nuclei, i.e., $E_n$ (2) the total number of baryons, or the ratio electric and gravitational forces, i.e., $N_B$ (3) the density parameter of the universe $\Omega$ (4) the amplitude of density fluctuations in the expanding universe (5) 'n'- the number of special dimensions.

The sixth is of course the cosmic vacuum term $\Lambda$.

It is usually thought (including by Rees himself) [15] that all these numbers are independent and not linked. One cannot predict one from another. However, one would expect the dominant cosmic vacuum energy (constituting most of the universe) to play some role in determining the other parameters. We shall explore this aspect, i.e., the cosmological constant as a unifying link.

Given a $\Lambda$ dominated universe, the requirement that for various large scale structures (held together by self gravity) to form a variety of length scales, their gravitational self energy density should at least match the ambient vacuum energy repulsion, was shown to imply [16, 17] a scale invariant mass-radius relationship to the form (for the various structures):

$$\frac{M}{R^2} \approx \sqrt{\Lambda}\frac{c^2}{G} \qquad \ldots(1)$$

Equation (1) predicts a universality of $M/R^2$ for a large variety of structures. For a typical spiral galaxy, $M_{gal} \approx 10^{12} M_\Theta, R \approx 30 kpc$, for globular clusters, $M \approx 10^6 M_\Theta, R \approx 100 pc$, for galaxy clusters, $M_C \approx 10^{16} M_\Theta, R_C \approx 3 Mpc$, for these and other structures, equation (1) holds! This equation can be easily shown to imply rough equality of $\rho_\Lambda, \rho_m$, etc. [18]

Let us now see whether $\Lambda$ and $E_n$ (nuclear binding energy) could be linked. For a nucleus of mass number A and radius r, the binding surface energy can be written as: $4\pi r^2 \left(A^{2/3} - 1\right)T$, where, T is the 'surface tension' of the nuclear force, i.e., energy per unit area (nucleus behaves like a liquid drop). For the helium nucleus $A = 4 \Rightarrow A^{2/3} \approx 2.5$. So the nuclear binding energy (for helium) becomes, $\Delta E_n = 6\pi r^2 T$. Now in [16-18] it was noted that remarkably enough T, which is essentially the energy per unit area, is just the same as given in equation (1) (the underlying physics was explored) (indeed $c^4/G$ is the superstring tension!)

$$T = \frac{\sqrt{\Lambda}}{G}c^4 \approx 10^{20} \, ergs/cm^2 \qquad \ldots(2)$$

This gives (when substituted into $\Delta E_n$) for the binding energy of the helium nucleus as, $\Delta E_n = 4.5 \times 10^{-5} \, ergs$, that is,

$$4\pi \left(\frac{\hbar}{m_\Pi c^2}\right)\left(A^{2/3} - 1\right)\frac{c^4}{G}\sqrt{\Lambda} = \Delta E_n = 4.5 \times 10^{-5} \, ergs \qquad \ldots(3)$$

This gives $E_n$ as 0.007 of the rest energy. So equation (1) not only gives the surface energy of large-scale structures (galaxies, globular clusters, superclusters, etc.) but also seems to fix the nuclear surface tension T, giving the Rees number $E_n$ as 0.007. There have been recent attempts to understand the coincidence between the vacuum energy $\Lambda$ and the matter density $\rho_m$. One has $\rho_\Lambda^{1/4} = \frac{M_W^2}{M_{pl}}$, where $M_W$ is the electroweak scale. This is very similar to our earlier work [19] where a $\Lambda$ term of the observed value was obtained through the electroweak vacuum made up of the weak boson condensate:

$$\Lambda \approx 8\pi G \frac{M_W^7}{M_{pl}^3} \approx 10^{-56} cm^{-2} \qquad \ldots(4)$$

In [19], $\Lambda$ was also related to the QCD strong interaction scale ($\Lambda_{QCD} \approx 160 MeV$) so we had the beautiful relation:

$$M_W^7 = M_{pl} \Lambda_{QCD}^6 \qquad \ldots(5)$$

($M_{pl}$ being the Planck scale)

Masses of particles like the electron $(m_e)$ and the fine structure constant $(\alpha)$ that have remained constant during the evolution of the universe could also be related to the cosmic vacuum energy (also constant!). Considering a wave packet of spread r, matching its gravitational self energy density to the cosmic repulsive vacuum density gives the required size of the wave packet and if it is charged, then this gives its mass as:

$$m_e = \frac{\alpha \hbar}{c} \Lambda_{pl}^{1/3} \Lambda^{1/6} \qquad \ldots(6)$$

It precisely turns out to be the electron mass! No a priori reason to expect this. $\Lambda_{pl} = \frac{c^3}{\hbar G}$. Note the weak dependence on $\Lambda$. If $G_F$ is the Fermi constant:

$\alpha = \Lambda_{pl}^{2/3} \Lambda^{1/6} \left(\frac{G_F}{\hbar c}\right)^{1/2}$. Similar formula [19] for the gluon coupling QCD, gives $\alpha_S \approx 0.13$, close to the low energy measured lab value! Similar formula for proton mass. Several intriguing relations of this sort are given in [20]. Again in [21], the baryon density was related to the dark (cosmic vacuum) energy as:

$$\frac{\rho_B}{\rho_C} \approx \frac{\sigma_T c^2 \Lambda^{1/2}}{\pi^2 G m_P} \approx 0.05 \qquad \ldots(7)$$

Equations (6) and (7), now relate the numbers $N_B$ and $\Omega$ of Rees. The ratio between electric and gravitational force is shown to be $1/(r_B \sqrt{\Lambda})$, where, $r_B$ is the Bohr radius.

Assuming DM to consist of collisionless particles (with velocity $v_d$) just bound by their self-gravity, the ratio of DM density $\rho_d$ to DE density is shown to be [22]:

$$\frac{\rho_d}{\rho_\Lambda} \approx \frac{2v_d^2}{c^2}\frac{1}{\Lambda r^2} \qquad \ldots(8)$$

For the largest structures, $r \approx 200 Mpc, v \approx 2000 km/s$ (largest dispersion velocities), this gives $\frac{\rho_d}{\rho_\Lambda} \approx 1/3$ as is observed! Again, the amplitude of the density fluctuations (the number Q of Rees) for large-scale structures (using equation (8)):

$$\Delta\phi \approx \frac{GM}{Rc^2} \approx \sqrt{\Lambda}R \approx 10^{-4} \qquad \ldots(9)$$

More exact derivation [22] gives $Q \approx 10^{-5}$. All of the above relations would be consistent with each other only for three special dimensions. In short an attempt has been made to understand how the various cosmological parameters acquire their present values and the probable seminal role of the cosmological constant in fixing the coupling strengths of the interactions and the masses of elementary particles.

There is strong recent evidence for the constancy of both the fundamental constants and DE, which dominates the universe. This approach also connects all the numbers of Rees, the cosmological constant (dominating the universe) playing the role of a unifying link.

**Reference:**

1. S Perlmutter, et al, Ap. J, 517, 46, 1999
2. Supernova Search Team Collab. (J P Blakes Lee, et al), Ap. J, 589, 693, 2003
3. A G Reiss, et al, Ap. J, 560, 49, 2001
4. de Bernardis, et al, Nature, 404, 955, 2001
5. WMAP Collab. (D Spergel, et al) Ap. J Suppl. Ser., 148, 175, 2003
6. A G Reiss, et al, Astro-ph/001384
7. S Carroll, et al, Astro-ph/0004075
8. S Allen, Mon. Not. Roy. Astron. Soc. (in press), 2006
9. L M Krauss, Ap. J, 604, 481, 2004
10. For example, S and T, 23, April 2006
11. K K Webb, et al, PRL, 87, 091301, 2001
12. T P Asenfelter, et al, PRL, 92, 041102, 2004
13. C Sivaram, Int. J. Theor. Phys., 33, 2407, 1994
14. C Sivaram, Amer. J. Phys., 50, 279; Astr. Spc. Sci, 82, 507, 1982
15. M Rees, *Just Six Numbers*, CUP, 2000
16. C Sivaram, Astr. Spc. Sci, 219, 135; IJTP, 33, 2407, 1994
17. C Sivaram, Mod. Phys. Lett., 34, 2463, 1999
18. C Sivaram, Astr. Spc. Sci., 271, 321; 219, 135, 2000
19. C Sivaram, Ap. J, 520, 317; Mod. Phys. Lett., 34, 2463, 1999
20. C Sivaram, Astr. Spc. Sci., 271, 361, 2000
21. C Sivaram, 21st Century Astrophysics, Ed. S K Saha and Rastogi, pp. 16-26, 2006
22. C Sivaram, in preparation (work in progress), 2006